\documentclass[
reprint,
amsmath,
amssymb,
aps,
superscriptaddress,
notitlepage
]{revtex4-2}

\usepackage[colorlinks, breaklinks=true, linkcolor=blue, citecolor=blue, linktocpage=true]{hyperref}
\usepackage{color}
\usepackage{graphicx}
\usepackage{dcolumn}
\usepackage{bm}

\begin{document}

\title{Microwave-to-Optical Quantum Transduction Utilizing the Topological Faraday Effect of Topological Insulator Heterostructures}

\author{Akihiko Sekine}
\email{akihiko.sekine@fujitsu.com}
\affiliation{Fujitsu Research, Fujitsu Limited, Kawasaki 211-8588, Japan}
\author{Mari Ohfuchi}
\affiliation{Fujitsu Research, Fujitsu Limited, Kawasaki 211-8588, Japan}
\author{Yoshiyasu Doi}
\affiliation{Fujitsu Research, Fujitsu Limited, Kawasaki 211-8588, Japan}

\date{\today}

\begin{abstract}
The quantum transduction between microwave and optical photons is essential for realizing scalable quantum computers with superconducting qubits.
Due to the large frequency difference between microwave and optical ranges, the transduction needs to be done via intermediate bosonic modes or nonlinear processes.
So far, the transduction efficiency $\eta$ via the magneto-optic Faraday effect (i.e., the light-magnon interaction) in the ferromagnet YIG has been demonstrated to be small as $\eta\sim 10^{-8} \mathrm{-} 10^{-15}$ due to the weak magneto-optic coupling.
Here, we take advantage of the fact that three-dimensional topological insulator thin films exhibit a topological Faraday effect that is independent of the sample thickness in the terahertz regime.
This leads to a large Faraday rotation angle and therefore enhanced light-magnon interaction in the thin film limit.
We show theoretically that the transduction efficiency between microwave and terahertz photons can be greatly improved to $\eta\sim10^{-4}$ by utilizing the heterostructures consisting of topological insulator thin films such as Bi$_2$Se$_3$ and ferromagnetic insulator thin films such as YIG. 
\\
\\
\end{abstract}

\maketitle

\section{Introduction}
The quantum transduction, or quantum frequency conversion, is an important quantum technology which enables the interconnects between quantum devices such as quantum processors and quantum memories.
In particular, the quantum transduction between microwave and optical photons has so far gathered attention in pursuit of large-scale quantum computers with superconducting qubits \cite{Lauk2020,Lambert2020,Han2021}.
Due to the large frequency difference between microwave and optical ranges, the transduction needs to be done via intermediate interaction processes with bosonic modes or via nonlinear processes, such as the optomechanical effect \cite{Andrews2014,Vainsencher2016,Higginbotham2018,Forsch2020,Mirhosseini2020,Jiang2020,Han2020,Arnold2020,Hoenl2022,Barzanjeh2022} electro-optic effect \cite{Rueda2016,Fan2018,Hease2020,Holzgrafe2020,McKenna2020,Xu2021,Youssefi2021,Sahu2022}, and magneto-optic effect \cite{Hisatomi2016,Osada2016,Zhang2016,Haigh2016,Zhu2020}.
To date, the transduction efficiency $\eta$, whose maximum value is $1$ by definition, has recorded the highest value $\eta\sim 10^{-1}$ with a bandwidth $\sim 10^{-2}\, \mathrm{MHz}$ \cite{Higginbotham2018} ($\eta\sim 10^{-2}$ with $\sim 1\, \mathrm{MHz}$ \cite{Fan2018}) among the transductions utilizing the optomechanical effect (electro-optic effect).

The focus of this paper is the microwave-to-optical quantum transduction via the magneto-optic Faraday effect, i.e., the light-magnon interaction.
Such a quantum transduction mediated by ferromagnetic magnons can have a wide bandwidth $\sim 1\, \mathrm{MHz}$ and can be operated even at room temperature \cite{Lauk2020,Lambert2020,Han2021}.
Also, the coherent coupling between a ferromagnetic magnon and a superconducting qubit has been realized \cite{Tabuchi2015,Viennot2015,Rameshti2022}.
However, the current major bottleneck when using the ferromagnetic insulators (FIs) such as YIG is the low transduction efficiency $
\eta\sim 10^{-8} \mathrm{-} 10^{-15}$ \cite{Hisatomi2016,Osada2016,Zhang2016,Haigh2016,Zhu2020,Wolf2007} due to the small light-magnon interaction strength $\zeta$.
The purpose of this study is to challenge this issue by utilizing topological materials, which are a new class of materials that are expected to exhibit unusual materials properties due to their topological nature.

In this paper, we take advantage of the fact that three-dimensional (3D) topological insulator (TI) thin films exhibit a topological Faraday effect that is independent of the sample thickness in the terahertz regime, leading to a large Faraday rotation angle and thus enhanced light-magnon interaction strength $\zeta$ in the thin film limit.
There has been a lot of works on improving the magneto-optic coupling (i.e., the Faraday effect) in ferromagnets.
While reducing the sample size has already been proposed to enhance the Faraday effect in earlier studies (see, for example, Ref.~\cite{Rameshti2022}), our main proposal of utilizing the topological Faraday effect in topological insulators has not been discussed.
To this end, we particularly consider the heterostructures consisting of TI thin films such as Bi$_2$Se$_3$ and FI thin films such as YIG.
We find that the transduction efficiency $\eta$ is inversely proportional to the thickness of the FI layers $d_{\mathrm{FI}}$, i.e., $\eta\sim\zeta\propto 1/d_{\mathrm{FI}}$, which is in sharp contrast to the  case of conventional FIs ($\eta\sim\zeta\propto d_{\mathrm{FI}}$).
We show theoretically that the transduction efficiency can be greatly improved to $\eta\sim 10^{-4}$ in a heterostructure of a few dozen of layers of nanometer-thick TI and FI thin films.

\section{Quantum transduction}
In this section, we describe a theory for the quantum transduction between microwave and terahertz photons in the TI heterostructures.
Particularly, we derive explicit expressions for the microwave-magnon and light-magnon interaction strengths in the TI heterostructures, which are the two important ingredients to evaluate the transduction efficiency.

\subsection{General consideration}
\begin{figure}[!t]
\centering
\includegraphics[width=\columnwidth]{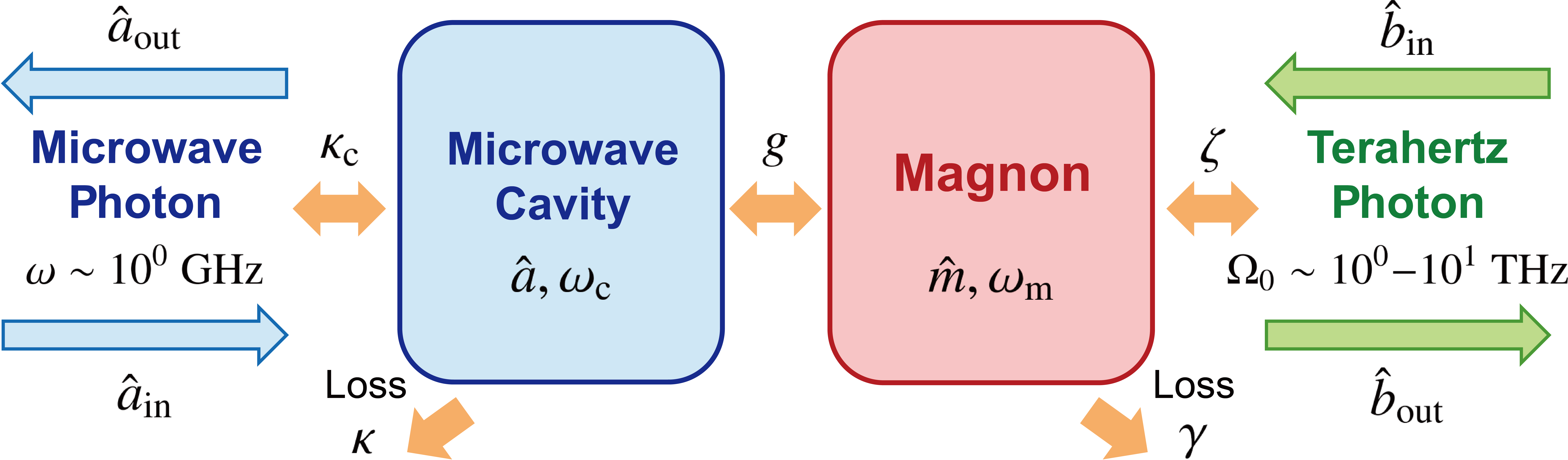}
\caption{Schematic illustration of our setup in terms of the operators ($\hat{a}_{\mathrm{in}}$, $\hat{a}_{\mathrm{out}}$, $\hat{b}_{\mathrm{in}}$, $\hat{b}_{\mathrm{out}}$, $\hat{a}$, and $\hat{m}$), coupling strengths ($\kappa_{\mathrm{c}}$, $g$, and $\zeta$), and losses ($\kappa$ and $\gamma$).
}\label{Fig1}
\end{figure}

Let us start with a generic description of our setup depicted in Fig.~\ref{Fig1}.
We consider the interaction Hamiltonian $H_{\mathrm{int}}=H_\kappa+H_g+H_\zeta$ \cite{Hisatomi2016}, where
\begin{align}
H_\kappa=-i\hbar\sqrt{\kappa_{\mathrm{c}}}\int_{-\infty}^\infty\frac{d\omega}{2\pi}\left[\hat{a}^\dag \hat{a}_{\mathrm{in}}(\omega)-\hat{a}_{\mathrm{in}}^\dag(\omega)\hat{a}\right]
\end{align}
describes the coupling between the microwave cavity photon $\hat{a}$ and an itinerant microwave photon $\hat{a}_{\mathrm{in}}(\omega)$,
\begin{align}
H_g=\hbar g\left(\hat{a}^\dag \hat{m}+\hat{m}^\dag \hat{a}\right)
\label{H_g}
\end{align}
describes the coupling between the microwave cavity photon and the magnon (in the ferromagnetic resonance state) $\hat{m}$, and
\begin{align}
&H_\zeta=-i\hbar\sqrt{\zeta}\nonumber\\
&\times\int_{-\infty}^\infty\frac{d\Omega}{2\pi}\left(\hat{m}+\hat{m}^\dag\right)\left[\hat{b}_{\mathrm{in}}(\Omega)e^{i\Omega_0 t}-\hat{b}^\dag_{\mathrm{in}}(\Omega)e^{-i\Omega_0 t}\right]
\label{H_zeta}
\end{align}
describes the coupling between the magnon and an itinerant optical photon $\hat{b}_{\mathrm{in}}(\Omega)$, which is indeed the sum of the beam-splitter-type and parametric-amplification-type interactions \cite{Hammerer2010}.
$\Omega_0$ is the input light frequency.
In order to relate the incoming and outgoing itinerant photons, we can employ the standard input-output formalism, to obtain
$\hat{a}_{\mathrm{out}}=\hat{a}_{\mathrm{in}}+\sqrt{\kappa_{\mathrm{c}}}\hat{a}$ and
$\hat{b}_{\mathrm{out}}=\hat{b}_{\mathrm{in}}+\sqrt{\zeta}\hat{m}$.
We solve the equations of motion for the cavity and magnon modes in the presence of intrinsic losses $\kappa$ and $\gamma$:
\begin{align}
\dot{\hat{a}}&=\frac{i}{\hbar}\left[H_{\mathrm{total}},\hat{a}\right]-\frac{\kappa_{\mathrm{c}}+\kappa}{2}\hat{a}-\sqrt{\kappa_{\mathrm{c}}}\hat{a}_{\mathrm{in}},\\
\dot{\hat{m}}&=\frac{i}{\hbar}\left[H_{\mathrm{total}},\hat{m}\right]-\frac{\gamma}{2}\hat{m}-\sqrt{\zeta}\hat{b}_{\mathrm{in}},
\end{align}
where $H_{\mathrm{total}}=H_0+H_{\mathrm{int}}$ is the total Hamiltonian of the system with $H_0$ being the noninteracting Hamiltonian for the cavity and magnon modes.

The microwave-to-optical quantum transduction efficiency, which is defined by the ratio between the outgoing and incoming photon numbers $\eta(\omega)=\left|\frac{\langle\hat{a}_{\mathrm{out}}(\omega)\rangle}{\langle\hat{b}_{\mathrm{in}}(\Omega)\rangle}\right|^2=
\left|\frac{\langle\hat{b}_{\mathrm{out}}(\Omega)\rangle}{\langle\hat{a}_{\mathrm{in}}(\omega)\rangle}\right|^2$, is obtained as \cite{Hisatomi2016}
\begin{align}
\eta(\omega)=\frac{4\mathcal{C}\frac{\kappa_{\mathrm{c}}}{\kappa_{\mathrm{c}}+\kappa}\frac{\zeta}{\gamma}}{\left(\mathcal{C}+1-4\frac{\Delta_{\mathrm{c}}}{\kappa_{\mathrm{c}}+\kappa}\frac{\Delta_{\mathrm{m}}}{\gamma}\right)^2+4\left(\frac{\Delta_{\mathrm{c}}}{\kappa_{\mathrm{c}}+\kappa}+\frac{\Delta_{\mathrm{m}}}{\gamma}\right)^2},
\label{Transduction-efficiency}
\end{align}
where $\mathcal{C}=\frac{4g^2}{(\kappa_{\mathrm{c}}+\kappa)\gamma}$ is the cooperativity, $\Delta_{\mathrm{c}}=\omega-\omega_{\mathrm{c}}$ is the detuning from the microwave cavity frequency $\omega_{\mathrm{c}}$, and $\Delta_{\mathrm{m}}=\omega-\omega_{\mathrm{m}}$ is the detuning from the ferromagnetic resonance frequency $\omega_{\mathrm{m}}$.
In this work we extend the above formalism to the magnetic heterostructures, deriving explicitly the expressions for $g$ and $\zeta$.

\subsection{Topological insulator heterostructures}
It has been shown that the magnitude of the magnon-mediated transduction efficiency~(\ref{Transduction-efficiency}) is essentially determined by the light-magnon coupling strength $\zeta$, i.e., $\eta\propto \zeta \propto \phi_{\mathrm F}^2/N_{\mathrm{s}}$ \cite{Hisatomi2016}, where $\phi_{\mathrm F}$ and $N_{\mathrm{s}}$ are respectively the Faraday rotation angle and number of net spins of the ferromagnet.
From this relation we see that the transduction efficiency can be improved in materials which exhibit a large Faraday rotation angle even with a small sample size.

We take advantage of the fact that 3D TI thin films exhibit a topological Faraday effect arising from the surface anomalous Hall effect, whose rotation angle $\phi_{\mathrm{F,TI}}$ is independent of the material thickness \cite{Tse2010,Maciejko2010,Tse2011,Sekine2021}.
Here, the bandgap $2\Delta$ of the surface Dirac bands, generated by the exchange coupling between the surface electrons and the proximitized magnetic moments having the components perpendicular to the surface, is essential for the occurrence of the surface anomalous Hall effect.
In other words, $\phi_{\mathrm{F,TI}}=0$ in the absence of proximitized magnetic moments.
The applicable range of the input light frequency $\Omega_0$ is limited by the cutoff energy $\varepsilon_{\mathrm{c}}$ of the surface Dirac bands (given typically by the half of the TI bulk bandgap), such that $\hbar\Omega_0 < \varepsilon_{\mathrm{c}}$ \cite{Tse2010}.
In particular, $\phi_{\mathrm{F,TI}}$ takes a universal value in the low-frequency limit $\hbar\Omega_0 \ll \varepsilon_{\mathrm{c}}$ and when the Fermi level $\mu_{\mathrm{F}}$ is in the bandgap $2\Delta$ \cite{Tse2010,Maciejko2010,Tse2011}
\begin{align}
\phi_{\mathrm{F,TI}}=\tan^{-1}\alpha\approx \alpha,
\label{Topological-Faraday-effect}
\end{align}
where $\alpha=e^2/\hbar c\approx 1/137$ is the fine-structure constant.
This universal behavior has been experimentally observed  \cite{Wu2016,Okada2016,Dziom2017,Sekine2021}.

\begin{figure}[!t]
\centering
\includegraphics[width=\columnwidth]{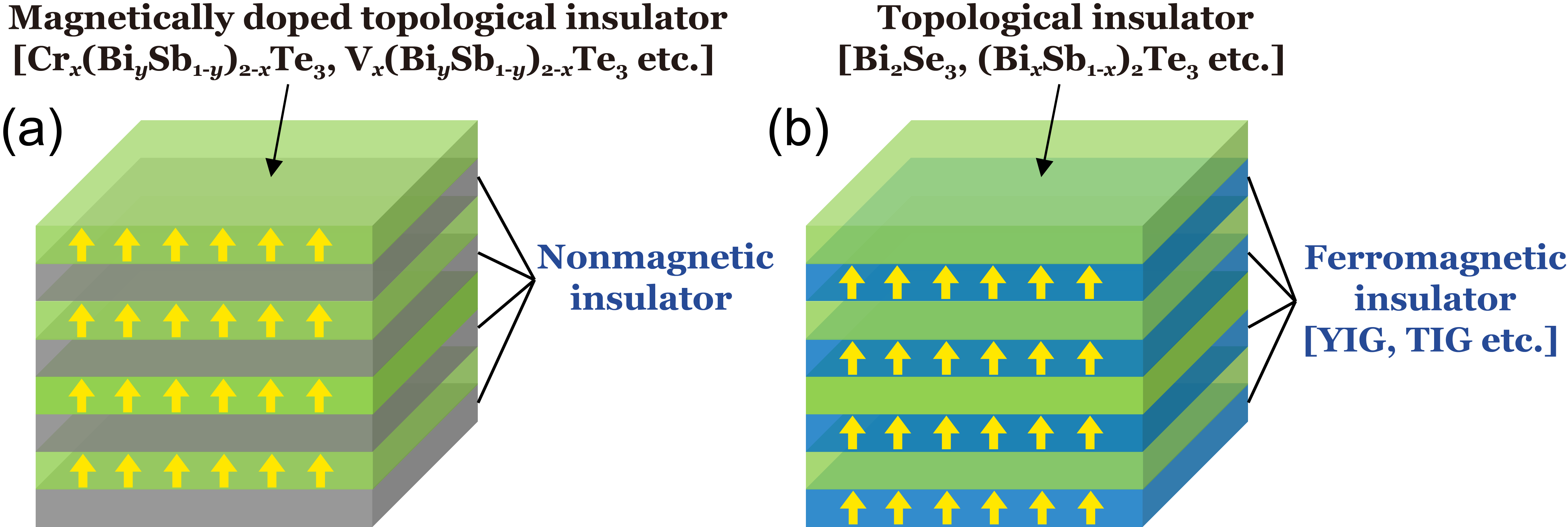}
\caption{(a) Heterostructure consisting of a magnetically doped TI and a nonmagnetic insulator.
(b) Heterostructure consisting of a (nonmagnetic) TI and a FI.
}\label{Fig2}
\end{figure}
We propose to utilize two types of TI heterostructures \cite{Tokura2019,Bhattacharyya2021,Liu2023}, as shown in Fig.~\ref{Fig2}.
One is the heterostructures consisting of magnetically doped TIs and nonmagnetic insulators \cite{Tokura2019,Liu2023}.
The other is the heterostructures consisting of (nonmagnetic) TIs and FIs \cite{Bhattacharyya2021,Liu2023,Tang2017,Fanchiang2018,Watanabe2019}.
In what follows, we focus on the latter because the surface anomalous Hall effect (and thereby the quantum transduction) can occur at a higher temperature $\sim 100\, \mathrm{K}$ than the former \cite{Tang2017,Fanchiang2018}.

\subsection{Light-magnon interaction in the TI heterostructure}
\begin{figure}[!t]
\centering
\includegraphics[width=\columnwidth]{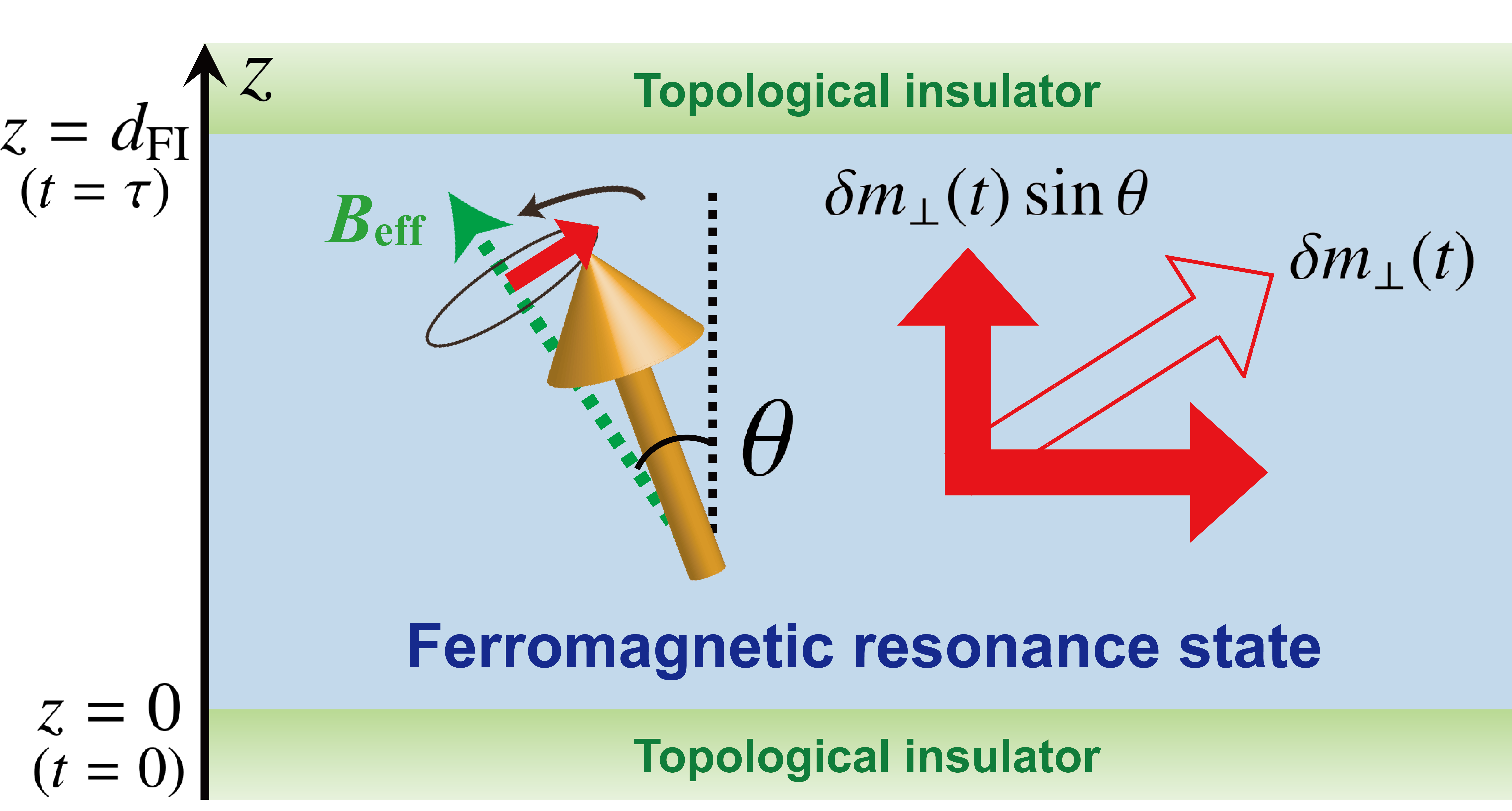}
\caption{Enlarged view of a heterostructure consisting of TIs and FIs.
$\delta m_\perp(t)$ is the small precessing component around the direction of the effective field $\bm{B}_{\mathrm{eff}}$.
The applied magnetic field needs to be tilted from the $z$ axis in order to induce a finite angle $\theta$ between the $z$ axis and $\bm{B}_{\mathrm{eff}}$.
}\label{Fig3}
\end{figure}

Suppose that a linearly polarized light is propagating along the $z$ direction.
Microscopically, the Hamiltonian for the Faraday effect in a ferromagnet is described by the coupling between the $z$-component of the magnetization density and the $z$-component of the Stokes operator of the light \cite{Hammerer2010,Hisatomi2016}.
We extend this Hamiltonian to the heterostructure of $N_{\mathrm{L}}$ TI layers and $N_{\mathrm{L}}$ FI layers [see Figs.~\ref{Fig2}(b) and \ref{Fig3}] as
\begin{align}
H_{\mathrm{F}}=\hbar A \sum_{i=1}^{N_{\mathrm{L}}}\int_{t_i}^{t_i+\tau} dt \, G_i(t)m_{i,z}(t) \mathcal{S}_z(t),
\label{Hamiltonian-heterostructure}
\end{align}
where $i$ denotes the $i$-th FI layer, $ G_i(t)$ is the coupling constant, $A$ is the cross section of the light beam, and $\tau=d_{\mathrm{FI}}/c$ (with $d_{\mathrm{FI}}$ the thickness of each FI layer and $c$ the speed of light in the material) is the interaction time.
We assume that the coupling constant describing the topological Faraday effect [Eq.~(\ref{Topological-Faraday-effect})] takes a $\delta$-function form, since it is a surface effect.
Taking also into account the conventional contribution to the Faraday effect, $cG_0$,  which takes a constant value across the sample \cite{Hisatomi2016}, we obtain
\begin{align}
G_i(t)=cG_0+\tfrac{1}{2}G_{\mathrm{TI}}\delta(t-t_i)+\tfrac{1}{2}G_{\mathrm{TI}}\delta(t-t_i-\tau).
\label{Faraday-coupling-constant}
\end{align}
The $z$ component of the magnetization density $m_{i,z}(t)$ is given by \cite{Rameshti2022}
\begin{align}
m_{i,z}(t)=\delta m_\perp(t)\sin\theta=\frac{\sqrt{N_{\mathrm{s}}}}{2V}\sin\theta\left[\hat{m}_i(t)+\hat{m}_i^\dag(t)\right],
\label{Magnetization-density}
\end{align}
where $V$ ($N_{\mathrm{s}}$) is the volume (number of net spins) of each FI layer, and $\hat{m}_i(t)$ is the magnon annihilation operator in the $i$-th FI layer satisfying $[\hat{m}_i(t),\hat{m}_j^\dag(t)]=\delta_{ij}$.
Note that a finite angle $\theta$ between the $z$ axis and the effective field $\bm{B}_{\mathrm{eff}}=-\partial F/\partial \bm{m}_i$ (with $F$ the free energy of each FI layer), which can be realized by a tilt of the applied magnetic field from the $z$ axis, is required.
Here, let us define a collective magnon operator $\hat{m}(t)\equiv \frac{1}{\sqrt{N_{\mathrm{L}}}}\sum_{i=1}^{N_{\mathrm{L}}}\hat{m}_i(t)$ satisfying $[\hat{m}(t),\hat{m}^\dag(t)]=1$, in a similar way as spin ensembles \cite{Wesenberg2009,Imamoglu2009}.
Then, Eq.~(\ref{Hamiltonian-heterostructure}) is simplified to be
\begin{align}
H_{\mathrm{F}}=&\ \hbar A \sqrt{N_{\mathrm{L}}}\int_0^\tau dt \, \left[cG_0+\tfrac{1}{2}G_{\mathrm{TI}}\delta(t)+\tfrac{1}{2}G_{\mathrm{TI}}\delta(t-\tau)\right]\nonumber\\
&\times m_z(t) \mathcal{S}_z(t),
\label{Hamiltonian-heterostructure-simplified}
\end{align}
where $m_z(t)=\frac{\sqrt{N_{\mathrm{s}}}}{2V}\sin\theta[\hat{m}(t)+\hat{m}^\dag(t)]$.
See Appendix~\ref{Appendix-Light-magnon-interaction} for an alternate derivation of the light-magnon interaction in the TI heterostructures.
The $z$ component of the Stokes operator for the polarization of light $\mathcal{S}_z(t)$ is given by \cite{Hammerer2010}
\begin{align}
\mathcal{S}_z(t)=\frac{1}{2A}\left[\hat{b}_R^\dag(t)\hat{b}_R(t)-\hat{b}_L^\dag(t)\hat{b}_L(t)\right],
\label{Stokes-operator}
\end{align}
where $\hat{b}_R(t)$ [$\hat{b}_L(t)$] is the annihilation operator of the mode of the right-circular (left-circular) polarized light propagating in the $z$ direction.
For a strong $x$-polarized light we have $\hat{b}_{R,L}(t)=\frac{1}{\sqrt{2}}(\hat{b}_x\pm i\hat{b}_y)\simeq \frac{1}{\sqrt{2}}(\langle\hat{b}_x\rangle\pm i\hat{b}_y)$ \cite{Hammerer2010}, where $\langle\hat{b}_x\rangle=\sqrt{\frac{P_0}{\hbar\Omega_0}}e^{-i\Omega_0 t}$ with $P_0$ ($\Omega_0$) the power (angular frequency) of the input light.

Because the interaction time $\tau=d_{\mathrm{FI}}/c\sim 10^{-8}\, \textrm{m}/(3\times 10^{8}\, \textrm{m/s})\sim 10^{-17}\, \textrm{s}$ is much shorter than the time scale of the magnon dynamics (i.e., the ferromagnetic resonance frequency) $1/\omega_m\sim 10^{-10}\, \textrm{s}$, the operators in Eq.~(\ref{Hamiltonian-heterostructure-simplified}) can be regarded as constant during the interaction.
Then, setting $\hat{b}_y\equiv \hat{b}_{\mathrm{in}}$, we arrive at Eq.~(\ref{H_zeta}).
The light-magnon coupling strength $\zeta$ is obtained as
\begin{align}
\zeta=\frac{\phi_{\mathrm{F}}^2N_{\mathrm{L}}}{N_{\mathrm{s}}}\sin^2\theta\frac{P_0}{\hbar\Omega_0},
\label{Light-magnon-coupling}
\end{align}
where the Faraday rotation angle $\phi_{\mathrm{F}}=(G_0d_{\mathrm{FI}}+G_{\mathrm{TI}})n_{\mathrm{s}}/4$ with $n_{\mathrm{s}}=N_{\mathrm{s}}/V$ the net spin density.
As expected, this expression for $\phi_{\mathrm{F}}$ properly describes the physical situation: the conventional contribution is proportional to the thickness $d_{\mathrm{FI}}$, while the topological contribution is independent of the thickness, i.e., is a surface contribution.
The $N_{\mathrm{L}}$ dependence in Eq.~(\ref{Light-magnon-coupling}) may be understood from that the total Faraday rotation angle and total number of spins in the heterostructure are $N_{\mathrm{L}}\phi_{\mathrm{F}}$ and $N_{\mathrm{L}}N_{\mathrm{s}}$, respectively.

\subsection{Coupling between magnon and microwave-cavity photon}
Next, we calculate the coupling strength $g$ in the heterostructure.
The total Hamiltonian describing the coupling between magnon and microwave-cavity photon is given by the sum of the contribution from each FI layer,
$H_g=\sum_{i=1}^{N_{\mathrm{L}}}\hbar g_i(\hat{a}^\dag \hat{m}_i+\hat{m}^\dag_i \hat{a})$,
where $g_i=g_0\sqrt{N_{\mathrm{s}}}$ with $g_0$ the single-spin coupling strength \cite{Tabuchi2014,Zhang2014}.
As we have done in the case of the light-magnon coupling, we may introduce a collective magnon operator $\hat{m}(t)\equiv \frac{1}{\sqrt{N_{\mathrm{L}}}}\sum_{i=1}^{N_{\mathrm{L}}}\hat{m}_i(t)$ satisfying $[\hat{m}(t),\hat{m}^\dag(t)]=1$.
Then, we obtain Eq.~(\ref{H_g}) with the coupling strength
\begin{align}
g=g_0\sqrt{N_{\mathrm{L}}N_{\mathrm{s}}}.
\label{microwave-magnon}
\end{align}
The expression~(\ref{microwave-magnon}) may also be understood from that the total number of spins in the heterostructure is $N_{\mathrm{L}}N_{\mathrm{s}}$.

\subsection{Magnetization dynamics in the FI layer}
So far, we have treated the electronic response of the TI surface state, i.e., the topological Faraday effect [Eq.~(\ref{Topological-Faraday-effect})] as the magnonic response of the FI layer.
Indeed, these two pictures are equivalent because the effective spin model is derived by integrating out the electronic degree of freedom in the surface Dirac Hamiltonian coupled to the FI layer via the exchange interaction \cite{Wakatsuki2015} (see Appendix~\ref{Appendix-Effective-Spin-Model} for details).
The derived spin model takes the form of the exchange interaction and the easy-axis anisotropy when the chemical potential $\mu_{\mathrm{F}}$ lies in the mass gap of the surface Dirac fermions, i.e., $\mu_{\mathrm{F}}<|\Delta|$, while it takes the form of the Dzyaloshinskii--Moriya interaction when $\mu_{\mathrm{F}}>|\Delta|$ \cite{Wakatsuki2015}.

It has been shown that the rotation angle of the topological Faraday effect decreases as the carrier density, i.e., the value of $\mu_{\mathrm{F}}$, becomes larger \cite{Tse2010} (see Appendix~\ref{Appendix-Topological-Faraday-Effect} for details).
We point out that such a behavior is consistent with the above-mentioned effective spin model analysis.
Generally, the Faraday rotation angle is proportional to the net spin density as $\phi_{\mathrm{F}}\propto n_{\mathrm{s}}$.
On the other hand, as represented by the skyrmion lattice, a canted spin structure is favored due to the Dzyaloshinskii--Moriya interaction, which leads to the decrease of the magnetization density, i.e., the net spin density $n_{\mathrm{s}}$.
Therefore, the decrease of the value of $\phi_{\mathrm{F,TI}}$ from the case of $\mu_{\mathrm{F}}<|\Delta|$ to the case of $\mu_{\mathrm{F}}>|\Delta|$ can also be explained from the effective spin model analysis.

\section{Transduction efficiency}
We are now in a position to obtain the transduction efficiency $\eta$ in the TI heterostructures.
We use the typical (possible) values of YIG for the FI layer and those of microwave cavity: $n_{\mathrm{s}}= 2.1\times 10^{19}\mu_{\mathrm B}\, \mathrm{mm}^{-3}$, $g_0/2\pi= 40\, \mathrm{mHz}$, $\kappa/2\pi= 1\, \mathrm{MHz}$, $\kappa_{\mathrm{c}}/2\pi= 3\, \mathrm{MHz}$, $\gamma/2\pi= 1\, \mathrm{MHz}$ \cite{Hisatomi2016,Tabuchi2014}.
We assume the sample area size of $0.5\, \mathrm{mm}\times 0.5\, \mathrm{mm}$, and treat $N_{\mathrm{L}}$ and $d_{\mathrm{FI}}$ as key variables.
We also assume the topological transport regime of $\mu_{\mathrm{F}}<|\Delta|$.
In the following, we focus on the thin film regime for the FI layer where the conventional contribution to the Faraday rotation angle can be neglected as $\phi_{\mathrm{F,0}}=G_0 n_{\mathrm{s}}d_{\mathrm{FI}}/4\ll 1/137$, as well as the thin film regime for the TI layer where the thickness of the TI thin film can be neglected as $d_{\mathrm{TI}}\ll \lambda=2\pi c/\Omega_0$.

As can be seen from Eq.~(\ref{Transduction-efficiency}), the transduction efficiency $\eta$ is proportional to the light-magnon interaction strength $\zeta$.
Thus, we firstly consider the dimensionless prefactor $\eta/(\zeta/\gamma)$.
The cooperativity $\mathcal{C}$ is obtained as $\mathcal{C}=1.6\times 10^{-15}\times N_{\mathrm{L}}N_{\mathrm{s}}$.
Assuming $N_{\mathrm{L}}\sim 10^{1}$ and $d\sim 1\, \mathrm{nm}$, we find that $\mathcal{C}\sim 10^{-1}$.
Accordingly, it turns out that the prefactor $\eta/(\zeta/\gamma)$ takes a maximum value $\approx 0.5$ when $\Delta_c=\Delta_m=0$, i.e., $\omega=\omega_c=\omega_m$.
we note that even when the cooperativity $\mathcal{C}$ is as large as $\mathcal{O}(10^{2})$, the maximum value of the prefactor $\eta/(\zeta/\gamma)$ is $\approx 0.7$ at the nonzero detunings $\Delta_c\neq 0$ and $\Delta_m\neq 0$.
From this result we see that the magnitude of $\eta$ in our case is also essentially determined by $\zeta$.

\begin{figure}[!t]
\centering
\includegraphics[width=\columnwidth]{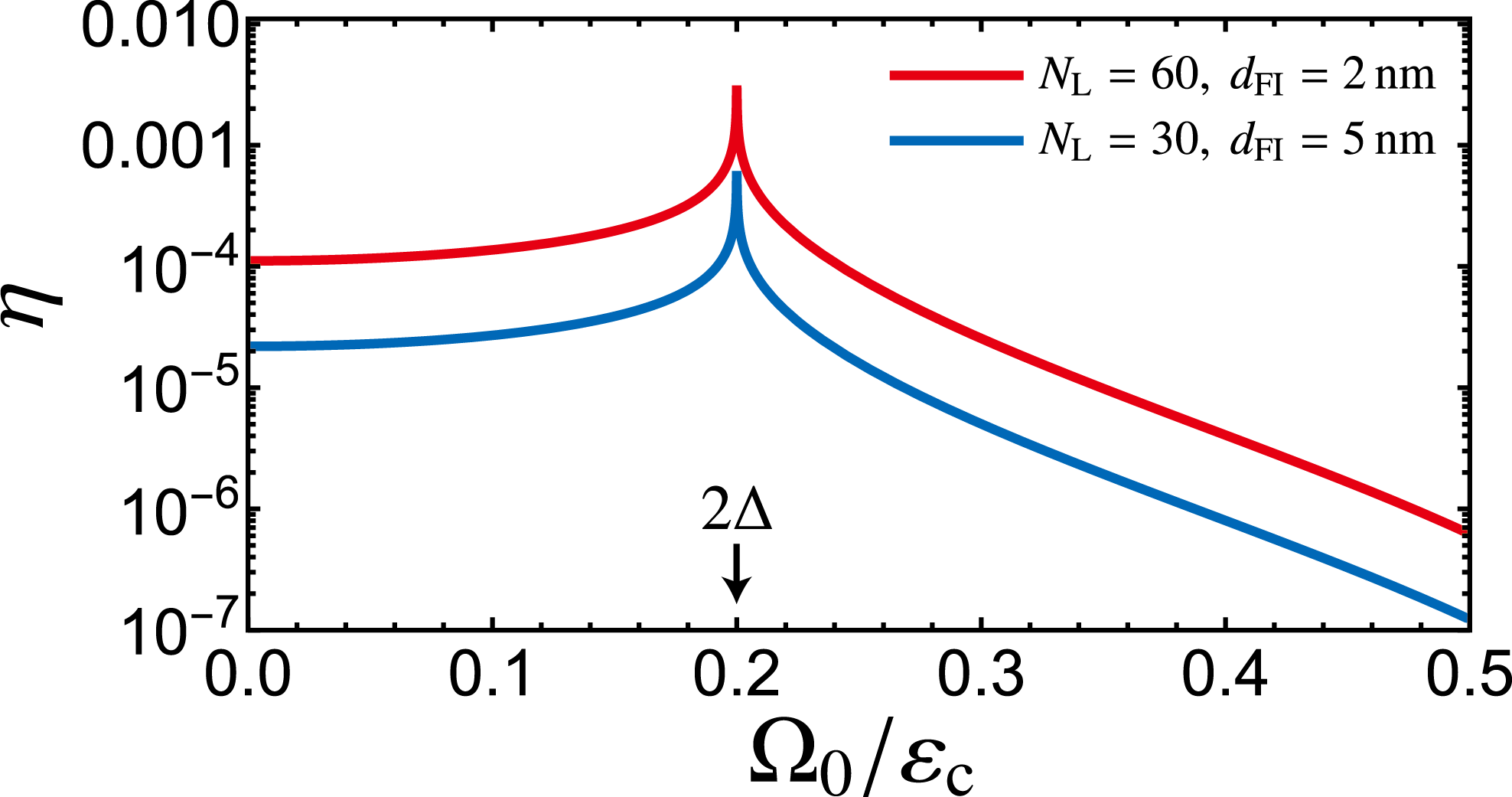}
\caption{The input light frequency $\Omega_0$ dependence of the transduction efficiency $\eta$.
We set $\theta=30^\circ$, $\varepsilon_{\mathrm{c}}=150\, \mathrm{meV}$, and $\Delta=15\, \mathrm{meV}$.
The input photon number is fixed here to $\frac{P_0}{\hbar\Omega_0}=1.5\times 10^{19}\, \mathrm{s^{-1}}$, which can be obtained, for example, with  $P_0=10\, \mathrm{mW}$ and $\Omega_0/2\pi=1\, \mathrm{THz}$.
}\label{Fig4}
\end{figure}

Next, we show the dependence of the transduction efficiency on the input light frequency $\Omega_0$ in Fig.~\ref{Fig4}.
Here, note that the input photon number $\frac{P_0}{\hbar\Omega_0}$ is fixed in Fig.~\ref{Fig4}.
In other words, the $\Omega_0$ dependence in Fig.~\ref{Fig4} originates from the topological Faraday rotation angle \cite{Tse2010} (see Appendix~\ref{Appendix-Topological-Faraday-Effect} for details).
Importantly, the Faraday rotation angle needs not be the universal value $\phi_{\mathrm{F,TI}}\approx 1/137$ (in the low frequency limit) and the transduction efficiency can be enhanced more than an order of magnitude near the sharp peak at the interband absorption threshold $2\Delta$ \cite{Tse2010} by tuning $\Omega_0$.
Assuming the cutoff energy $\varepsilon_{\mathrm{c}}=150\, \mathrm{meV}$ and the surface mass gap $\Delta=15\, \mathrm{meV}$ \cite{Tokura2019,Bhattacharyya2021,Liu2023}, we find that the frequency at which $\eta$ takes the maximum value is $\Omega_0/2\pi=7.3\, \mathrm{THz}$.
Note that the maximum value of $\eta$ is mainly determined by $N_{\mathrm{L}}$ and $d_{\mathrm{FI}}$, not by $\varepsilon_{\mathrm{c}}$ or $\Delta$.

If a heterostructure of $N_{\mathrm{L}}=500$ can be realized [where the total thickness of the heterostructure $2N_{\mathrm{L}}(d_{\mathrm{TI}}+d_{\mathrm{FI}})\approx 10\, \mathrm{\mu m}$ is still shorter than the wavelength $\lambda=2\pi c/\Omega_0$], the maximum value of $\eta$ in Fig.~\ref{Fig4} can be further improved to $\eta\sim 10^{-2}$, which is comparable to the current achievements in optomechanical and electro-optic systems \cite{Lauk2020,Lambert2020,Han2021}.
We note also that, by optimizing the size of YIG and utilizing the whispering gallery mode of the magnons of YIG sphere in an optical cavity, the microwave-to-optical transduction efficiency with YIG can theoretically reach $\sim 10^{-2}$ \cite{Osada2016,Zhang2016}.

The transduction efficiency $\eta\sim 10^{-3} \mathrm{-} 10^{-4}$ obtained in Fig.~\ref{Fig4} is greatly improved compared to that of a spherical YIG of $0.75\ \mathrm{mm}$ diameter, $\eta\sim 10^{-10}$, obtained with $P_0=15\, \mathrm{mW}$ and $\Omega_0/2\pi=200\, \mathrm{THz}$ \cite{Hisatomi2016}.
Here, it should be noted that the Verdet constant $\mathcal{V}(=G_0n_{\mathrm{s}}/4)$ in YIG in the terahertz range ($\sim 1\, $THz) is about an order of magnitude smaller than that in the telecom frequency range ($\approx 200\, $THz) \cite{Li2020}.
This means that the light-magnon interaction strength $\zeta$ does not change largely even in the terahertz range due to the relation $\zeta\propto \phi_{\mathrm{F,0}}^2/\Omega_0$ where $\phi_{\mathrm{F,0}}=\mathcal{V}d_{\mathrm{FI}}$.
Then, it follows that the transduction efficiency using YIG would be as small as $\mathcal{O}(10^{-10})$ even in the terahertz range \cite{Comment3}.

There is a fundamental difference in the sample size dependence of  the transduction efficiency $\eta$ between previous studies and our study, while the expression $\eta\sim\zeta\propto \phi_{\mathrm{F}}^2/N_{\mathrm{s}}$ is the same.
In conventional ferromagnets such as YIG, one obtains $\eta\propto d_{\mathrm{FI}}$ since both $\phi_{\mathrm{F}}$ and $N_{\mathrm{s}}$ are proportional to $d_{\mathrm{FI}}$.
This indicates that the value of $\eta$ becomes very small in the thin film limit.
On the other hand, in TI heterostructures, one obtains $\eta\propto 1/d_{\mathrm{FI}}$ since $\phi_{\mathrm{F}}$ is constant whereas $N_{\mathrm{s}}$ is proportional to $d_{\mathrm{FI}}$.
This is the mechanism for the enhancement of $\eta$ in the thin film limit.

While we have so far considered the case of the heterostructures consisting of a (nonmagnetic) TI and a FI, we can similarly treat the heterostructures consisting of a magnetically doped TI and a nonmagnetic insulator since the physics in both cases are the same.
See Appendix~\ref{Appendix-MTI} for an estimation of the transduction efficiency in the heterostructures consisting of a magnetically doped TI and a nonmagnetic insulator.

\begin{figure}[!t]
\centering
\includegraphics[width=\columnwidth]{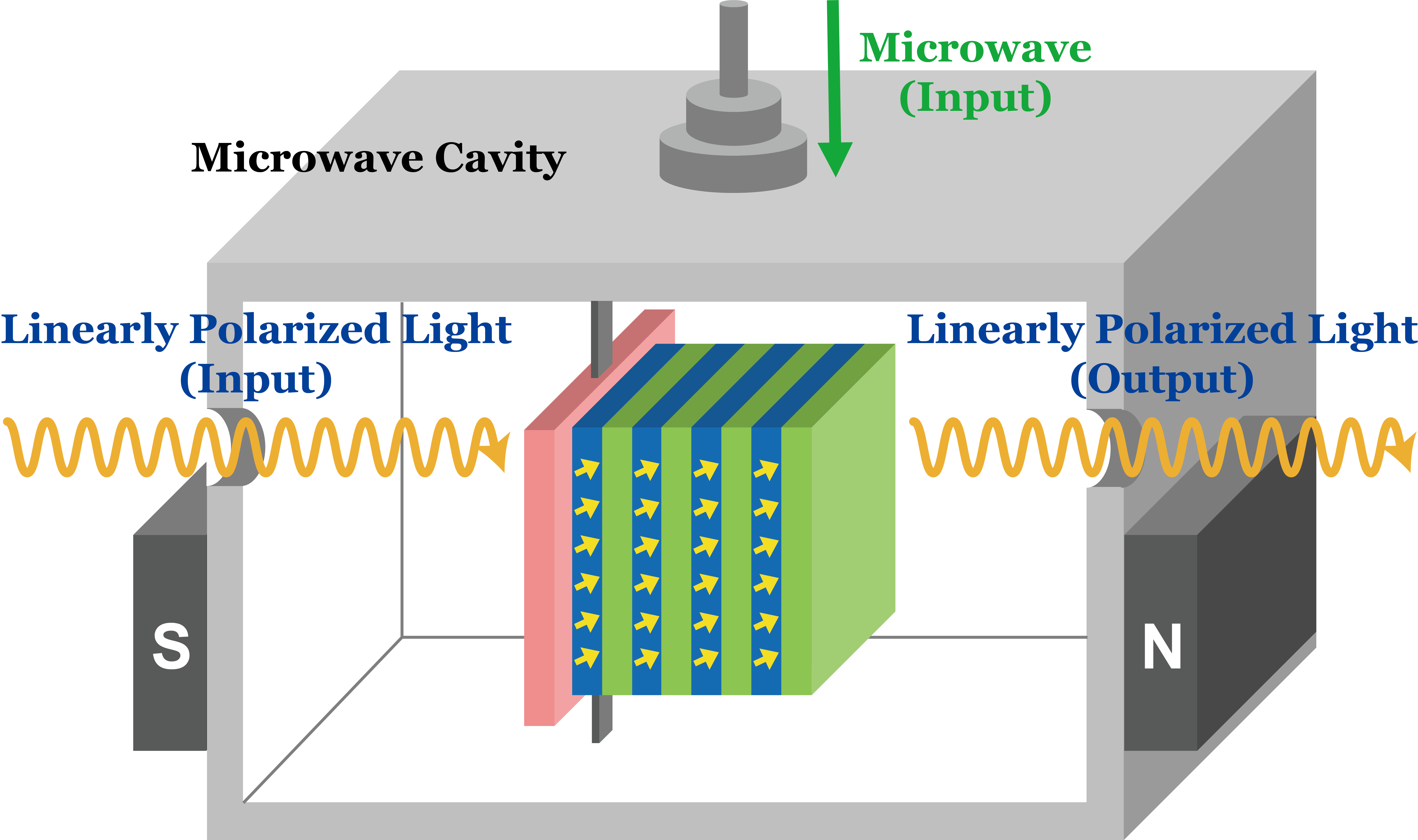}
\caption{
Schematic Illustration of the microwave-to-optical quantum transduction utilizing a TI heterostructure.
}\label{Fig5}
\end{figure}
Finally, we show in Fig.~\ref{Fig5} a schematic illustration of the microwave cavity setup in our microwave-to-optical quantum transduction.
Here, a linearly polarized light in the terahertz range is applied perpendicular to the heterostructure plane.
A finite tilt angle between the light propagation direction and the ground state direction of the ferromagnetic moments is required by applying a static magnetic field.

\section{Discussion and Summary}
We briefly discuss a possible application of our finding.
While optical fibers in the telecom frequency range are currently used widely, we would like to stress that optical fibers in the terahertz range are also in principle able to interconnect quantum devices.
Actually, terahertz optical fibers are under active research and development \cite{Islam2020}.
Thus, we expect that in the future TI heterostructures might be used as a quantum transducer for superconducting quantum computers interconnected via terahertz optical fibers.

One possible way for bringing the light frequency closer to the telecom frequency range is to find TIs with a large bulk bandgap ($\approx 2\varepsilon_{\mathrm{c}}$), as well as combinations of FIs and such TI surface states that allow a strong exchange interaction between them and therefore enable a large surface bandgap $2\Delta$.
If a TI heterostructure with $\Delta\approx 100\, \mathrm{meV}$ is discovered, then the maximum transduction efficiency will be obtained at $\Omega_0/2\pi=2\Delta\approx 48\, \mathrm{THz}$.
In this case, infrared optical fibers can be used.

We also expect that our finding can be useful in helping study the so-called terahertz gap \cite{Williams2006}.
For example, the readout of materials information (and inversely the manipulation of materials) in the terahertz gap may be possible via microwave.

To summarize, we have shown that the transduction efficiency of the microwave-to-optical quantum transduction mediated by ferromagnetic magnons can be greatly improved by utilizing the topological Faraday effect in 3D TI thin films.
By virtue of the topological contribution to the Faraday rotation angle which is independent of the thickness of the FI layer, the transduction efficiency is strongly enhanced in the thin film limit.
Our study opens up a way for possible applications of topological materials in future quantum interconnects.

\acknowledgements
We would like to thank Shintaro Sato, Ryo Murakami, Kenichi Kawaguchi, and Shoichi Shiba for their advice and support.

\appendix
\section{Alternate Derivation of the Light-Magnon Interaction \label{Appendix-Light-magnon-interaction}}
The light-magnon interaction~(\ref{Hamiltonian-heterostructure}) was originally derived as the Faraday effect in atomic ensembles in a magnetic field \cite{Hammerer2010}.
In this Appendix, following Refs.~\cite{Kusminskiy2016,Rameshti2022}, we show an alternative derivation of the light-magnon interaction in the TI heterostructures.
This derivation utilizes the fact that the permittivity tensor is a function of the magnetization $\bm{M}$ as $\hat{\varepsilon}_{ij}(\bm{M})$.
To linear order in the magnetization, we have \cite{LL-Book}
\begin{align}
\hat{\varepsilon}_{ij}(\bm{M})=\varepsilon_0\left(\varepsilon\delta_{ij}-if\sum_k\epsilon_{ijk}M_k\right),
\end{align}
where $\varepsilon_0$ ($\varepsilon$) is the vacuum (relative) permittivity, $\epsilon_{ijk}$ is the Levi-Civita symbol, and $f$ is a material-dependent constant.
The Hamiltonian describing the interaction between light and magnetization can be obtained from the electromagnetic energy $U=\int d^3r\, \bm{E}\cdot\bm{D}/2$, where $\bm{D}=\hat{\varepsilon} (\bm{M})\bm{E}$ with $\bm{E}$ an external electric field.
Then, the magnetization-dependent contribution is
\begin{align}
H_{\mathrm{int}}=-i\frac{\varepsilon_0}{4}\int d^3r\, f(\bm{r}) \bm{M}(\bm{r})\cdot\left[\bm{E}^*(\bm{r})\times\bm{E}(\bm{r})\right],
\label{magnetization-dependent-contribution}
\end{align}
where $\bm{E}$ and $\bm{E}^*$ are the real and imaginary parts of the complex electric field.
In Eq.~(\ref{magnetization-dependent-contribution}) we have introduced the spatial dependence in $f$, since the Faraday rotation angle has a spatial dependence in the TI heterostructures.

As we do in the main text, we assume that $f(\bm{r})$ takes a $\delta$-function form at the interface between the TI layer and the ferromagnetic insulator layer, while it takes a constant in the bulk of the $n$-th ferromagnetic insulator layer (which is denoted by $i$-th layer in the main text):
\begin{align}
f(\bm{r})=\tfrac{1}{2}f_{\mathrm{TI}}\delta(r_z-r_{z,n})+\tfrac{1}{2}f_{\mathrm{TI}}\delta(r_z-r_{z,n}-d_{\mathrm{TI}})+f_0.
\end{align}
The electric field can be quantized as $\bm{E}^+(\bm{r},t)=\sum_\alpha\bm{E}_\alpha(\bm{r})\hat{b}_\alpha(t)$ and $\bm{E}^-(\bm{r},t)=[\bm{E}^+(\bm{r},t)]^\dag$, where $\bm{E}_\alpha(\bm{r})$ is the $\alpha$-th eigenmode of the electric field and $\hat{b}_\alpha(t)$ is the annihilation operator of the photon in the $\alpha$-th eigenmode.
Because we are focusing on the ferromagnetic resonance state where the spins are precessing uniformly, we can set $\bm{M}(\bm{r})=\bm{M}$.
We also have the relation $M_i/M_{\mathrm{s}}=\hat{S}_i/S$ with $M_{\mathrm{s}}$ being the saturation magnetization and $S$ being the total number of spins (which scales as the sample volume).
Finally, we arrive at the interaction Hamiltonian
\begin{align}
H_{\mathrm{int}}=\hbar\sum_{i,\alpha,\beta}\mathcal{G}^i_{\alpha\beta}\hat{S}_i\hat{b}^\dag_\alpha\hat{b}_\beta,
\end{align}
where the coupling strength $\mathcal{G}^i_{\alpha\beta}$ is given by
\begin{widetext}
\begin{align}
\mathcal{G}^i_{\alpha\beta}=-i\frac{\varepsilon_0 M_{\mathrm{s}}}{4\hbar S}\sum_{j,k}\epsilon_{ijk}\int d^3r\,  \left[\tfrac{1}{2}f_{\mathrm{TI}}\delta(r_z-r_{z,n})+\tfrac{1}{2}f_{\mathrm{TI}}\delta(r_z-r_{z,n}-d_{\mathrm{TI}})+f_0\right]E^*_{\alpha,j}(\bm{r})E_{\beta,k}(\bm{r}).
\label{Coupling-constant-definition}
\end{align}
Here, note that the matrix $\mathcal{G}^i$ is Hermitian because the Hamiltonian $H_{\mathrm{int}}$ is Hermitian.
In general, the matrix $\mathcal{G}^i$ can be diagonalized in the basis of the optical polarization mode and the solution for the simplest case of $\mathcal{G}^i_{\alpha\alpha}\neq 0$ and $\mathcal{G}^i_{\alpha\beta}=0$ ($\alpha\neq \beta$) can be found \cite{Kusminskiy2016}.
In the case of a circularly polarized light propagating in the $z$ direction with the polarization basis $\bm{e}_{R,L}=\frac{1}{\sqrt{2}}(\bm{e}_x\pm i\bm{e}_y)$, we find that $\mathcal{G}^x=\mathcal{G}^y=0$ and
\begin{align}
\mathcal{G}\equiv \mathcal{G}^z_{RR}=-\mathcal{G}^z_{LL}=\frac{1}{S}\left[c\mathcal{G}_0+\tfrac{1}{2}\mathcal{G}_{\mathrm{TI}}\delta(t-t_n)+\tfrac{1}{2}\mathcal{G}_{\mathrm{TI}}\delta(t-t_n-\tau)\right],
\label{Coupling-constant}
\end{align}
where $\tau=d_{\mathrm{TI}}/c$.
The expression for $\mathcal{G}_0$ is related to the Faraday rotation angle per unit length $\theta_{\mathrm{F,0}}$ as $\mathcal{G}_0=\theta_{\mathrm{F,0}}\xi/(4\sqrt{\varepsilon})$ with $\xi$ a dimensionless constant \cite{Kusminskiy2016}.
Here, note that the dimension of $\mathcal{G}$ is $[\mathrm{time}]^{-1}$ \cite{Kusminskiy2016} and therefore we have explicitly written down the time dependence in the topological contribution [the last two terms in Eq.~(\ref{Coupling-constant})], which originates from the $\delta$ function in Eq.~(\ref{Coupling-constant-definition}) with the replacement $r_z\to c t$.
This replacement becomes valid when Eq.~(\ref{Coupling-constant}) is integrated over time, as we do in the main text.

As shown in Fig.~\ref{Fig3} in the main text, the ground-state magnetization direction is tilted from the $z$  axis with the angle $\theta$.
Thus, $\hat{S}_z$ is represented in terms of the small deviation of uniformly precessing spins in the ferromagnetic resonance state, $\delta m_\perp$, as $\hat{S}_z=\delta m_\perp\sin\theta\approx \sqrt{S/2}\sin\theta(\hat{m}+\hat{m}^\dag)$.
Also, in such cases where $\hat{b}$ and $\hat{b}^\dag$ represent itinerant photons (denoted by the subscript ``in'') as in the main text, the interaction Hamiltonian needs to be integrated over the interaction time.
(This can be checked by that the operators $\hat{b}_{\mathrm{in}}$ and $\hat{b}^\dag_{\mathrm{in}}$ have the dimension of $[\mathrm{time}]^{-1/2}$.)
Finally, we arrive at the light-magnon interaction for the $N_{\mathrm{L}}$-layer topological insulator heterostructure:
\begin{align}
H_{\mathrm{int}}=\frac{\hbar \sin\theta}{\sqrt{2S}}\sum_{i=1}^{N_{\mathrm{L}}}\int_{t_i}^{t_i+\tau} dt \, \left[c\mathcal{G}_0+\tfrac{1}{2}\mathcal{G}_{\mathrm{TI}}\delta(t-t_i)+\tfrac{1}{2}\mathcal{G}_{\mathrm{TI}}\delta(t-t_i-\tau)\right]\left(\hat{m}_i+\hat{m}^\dag_i\right)\left(\hat{b}^\dag_R\hat{b}_R-\hat{b}^\dag_L\hat{b}_L\right),
\end{align}
\end{widetext}
which is equivalent to Eq.~(\ref{Hamiltonian-heterostructure-simplified}) in the main text (apart from the numerical factor that can be absorbed by a redefinition of the coupling strengths).

\section{Magnetization Dynamics of the Ferromagnetic Insulator Layer \label{Appendix-Effective-Spin-Model}}
In this Appendix, we show that the effective spin model is derived by integrating out the electronic degree of freedom in the surface Dirac Hamiltonian coupled to the FI layer via the exchange interaction.
To this end we start with the surface-state Hamiltonian of the TI layer exchange-coupled to the FI layer, which is given by $H=\sum_{\bm{k}}\psi^\dag_{\bm{k}}\mathcal{H}(\bm{k})\psi_{\bm{k}}$ with 
\begin{align}
\mathcal{H}(\bm{k})&=\hbar v_{\mathrm{F}}(k_x\sigma_x+k_y\sigma_y)+J\bm{m}\cdot\bm{\sigma}\nonumber\\
&\equiv \hbar v_{\mathrm{F}}(k_x\sigma_x+k_y\sigma_y)+\Delta\sigma_z+\sum_\alpha J_\alpha\delta m_\alpha\sigma_\alpha,
\end{align}
where $v_{\mathrm{F}}$ is the Fermi velocity, $\sigma_i$ are the Pauli matrices describing electrons' spin on the TI surface, $J$ is the strength of the exchange coupling, and $\bm{m}=\bm{m}_0+\delta\bm{m}$ (with $\bm{m}_0$ the ground state direction) is the magnetization density vector of the FI layer.
The effective action for the FI layer can be derived by integrating out the electronic degree of freedom:
\begin{align}
Z&=\int\mathcal{D}[\psi,\bar{\psi}]\, e^{iS}\nonumber\\
&\equiv e^{iS_{\rm eff}[\delta \bm{m}]}=\exp\left[\mathrm{Tr}\left(\ln G_0^{-1}\right)-\sum_{n=1}^\infty\frac{1}{n}\mathrm{Tr}\left(-G_0V\right)^n\right],
\end{align}
where $G_0(\bm{k},i\omega_n)=[i\omega_n-\mathcal{H}(\bm{k})-\mu_{\mathrm{F}}]^{-1}$ is the unperturbed Green's function and $V=\sum_\alpha J_\alpha\delta m_\alpha\sigma_\alpha$  is a perturbation.
At the one-loop level and in the low-frequency limit \cite{Wakatsuki2015}, the effective action for $\delta \bm{m}$ is written in terms of the static susceptibility $\chi_{\alpha\beta}(\bm{q},0)$ as
\begin{align}
S_{\rm eff}[\delta\bm{m}]&=\frac{1}{2}i\mathrm{Tr}\left(G_0V\right)^2\nonumber\\
&=\frac{1}{2}\sum_{\bm{q}}\sum_{\alpha,\beta}J_\alpha J_\beta \delta m_\alpha(\bm{q})\chi_{\alpha\beta}(\bm{q},0)\delta m_\beta(-\bm{q}),
\end{align}
which takes the form of the exchange interaction and the easy-axis anisotropy when the chemical potential $\mu_{\mathrm{F}}$ lies in the mass gap of the surface Dirac fermions, i.e., $\mu_{\mathrm{F}}<|\Delta|$, while it takes the form of the Dzyaloshinskii--Moriya interaction when $\mu_{\mathrm{F}}>|\Delta|$  \cite{Wakatsuki2015}.
This indicates that the magnetization dynamics of the FI layer is modified by the presence of the TI surface state, depending on the value of the chemical potential.

\section{Topological Faraday Effect in Topological Insulator Thin Films \label{Appendix-Topological-Faraday-Effect}}
In this Appendix, following Refs.~\cite{Tse2010,Tse2011}, we derive an analytical expression for the Faraday rotation angle in a single TI surface.
Then, using the obtained expression, we show the input light frequency dependence and chemical potential dependence of the topological Faraday rotation angle in a TI thin film.

\subsection{Analytical expression for the Faraday rotation angle}
We consider the electronic response of the TI surface state which is described by  the effective Hamiltonian of the form,
\begin{align}
\mathcal{H}(\bm{k})=\hbar v_{\mathrm{F}}(k_x\sigma_x+k_y\sigma_y)+\Delta\sigma_z,
\label{Surface-Dirac-Hamiltonian}
\end{align}
where $v_{\mathrm{F}}$ is the Fermi velocity, $\sigma_i$ are the Pauli matrices describing electrons' spin on the TI surface, and $\Delta$ is the mass gap induced by the exchange coupling between the proximitized ferromagnetic moments.
The optical conductivity of the system described by Eq.~(\ref{Surface-Dirac-Hamiltonian}) can be obtained by solving a quantum kinetic equation \cite{Tse2010}.
In the following, the conductivities are given in units of $e^2/\hbar=c\alpha$, where we set $c=1$.
The longitudinal conductivity $\sigma_{xx}(\Omega_0)=\sigma_{xx}^{\mathrm{R}}(\Omega_0)+i\sigma_{xx}^{\mathrm{I}}(\Omega_0)$ and the transverse conductivity $\sigma_{xy}(\Omega_0)=\sigma_{xy}^{\mathrm{R}}(\Omega_0)+i\sigma_{xy}^{\mathrm{I}}(\Omega_0)$ at the frequency $\Omega_0$ of the incident electric field (i.e., light) are written explicitly as \cite{Tse2010}
\begin{align}
\sigma_{xx}^{\mathrm{R}}(\Omega_0)=&\ \frac{\mu_{\mathrm{F}} ^2-\Delta ^2}{4 \mu_{\mathrm{F}} }\, \delta (\Omega_0)\, \theta \left(\mu_{\mathrm{F}} -\Delta \right)\nonumber\\
&\ +\left[\left(\frac{\Delta }{2\Omega_0}\right)^2+\frac{1}{16}\right] \theta \left[\Omega_0 -2 \max (\mu_{\mathrm{F}} ,\Delta )\right],\\
\sigma_{xx}^{\mathrm{I}}(\Omega_0)=&\ \frac{1}{4 \pi}\frac{\mu_{\mathrm{F}} ^2-\Delta ^2}{\mu_{\mathrm{F}}  \Omega_0}\, \theta \left(\mu_{\mathrm{F}} -\Delta \right)\nonumber\\
&\ +\frac{1}{16\pi} \left[4 \left(\frac{\Delta }{\Omega_0}\right)^2+1\right] F(\Omega_0)\nonumber\\
&\ -\frac{1}{4\pi}\frac{\Delta ^2}{\Omega_0} \left(\frac{1}{\varepsilon _c}-\frac{1}{\max (\mu_{\mathrm{F}} ,\Delta )}\right),
\label{sigma_xx}
\end{align}
and
\begin{align}
\sigma_{xy}^{\mathrm{R}}(\Omega_0)&=-\frac{\Delta  }{4 \pi \Omega_0}F(\Omega_0),\\
\sigma_{xy}^{\mathrm{I}}(\Omega_0)&=\frac{\Delta }{4 \Omega_0}\, \theta \left[\Omega_0 -2 \max (\mu_{\mathrm{F}} ,\Delta )\right],
\label{sigma_xy}
\end{align}
where
\begin{align}
F(\Omega_0)=\ln \left| \frac{\Omega_0 +2 \varepsilon _c}{\Omega_0 -2 \varepsilon _c}\right| -\ln \left| \frac{\Omega_0 +2 \max (\mu_{\mathrm{F}} ,\Delta )}{\Omega_0 -2 \max (\mu_{\mathrm{F}} ,\Delta )}\right|,
\end{align}
$\varepsilon_{\mathrm{c}}$ is the cutoff energy of the surface Dirac bands $\pm\sqrt{\hbar^2v_{\mathrm{F}}^2(k_x^2+k_y^2)+\Delta^2}$ which is given typically by the half of the TI bulk bandgap, and we have considered the case of $\mu_{\mathrm{F}}>0$ and $\Delta>0$ without loss of generality.

In what follows, we derive an analytical expression for the Faraday rotation angle at a {\it single} interface. 
Suppose that an electromagnetic wave is propagating along the $z$ direction from medium $i$ to medium $j$ with dielectric constant $\epsilon_i$ and $\epsilon_j$ and magnetic permeability $\mu_i$ and $\mu_j$, respectively.
The boundary conditions for the electric and magnetic fields are \cite{Tse2011}
\begin{align}
\bm{E}^i=\bm{E}^j\ \ \mathrm{and}\ \ -i\tau_y(\bm{B}^j-\bm{B}^i)=\mu_0\bm{J}_{\mathrm{s}},
\label{Boundary-conditions}
\end{align}
where the first equation follows from the continuity of the electric field by Faraday's law and the second equation follows from Amp\`{e}re's law integrated over the $z$ direction $\int dz\, \nabla\times\bm{B}=\int dz\, \mu_0\bm{j}$.
$\tau_y$ is the $y$-component of the Pauli matrices.
Here, we have assumed that the bulk of the media is insulating, i.e., the electric current $\bm{J}_{\mathrm{s}}=\sigma\bm{E}^i$ flows only at the boundary.
Note also that $E_z=B_z=0$ because we are considering an transverse wave.

Let $\bm{E}^0$, $\bm{E}^r$, and $\bm{E}^t$ be incident, reflected, and transmitted electric fields, respectively.
Then, the electric fields in media $i$ and $j$ are given by
\begin{align}
\bm{E}^i&=e^{ik_iz}\bm{E}^{ti}+e^{-ik_iz}\bm{E}^{ri},\nonumber\\
\bm{E}^j&=e^{ik_jz}\bm{E}^{tj}.
\end{align}
At the boundary, the incoming fileds $\begin{bmatrix}\bm{E}^{ti} & \bm{E}^{rj}\end{bmatrix}^T$ and the outgoing fields $\begin{bmatrix}\bm{E}^{ri} & \bm{E}^{tj}\end{bmatrix}^T$ are related by the scattering matrix 
$S=\begin{bmatrix}\begin{smallmatrix}r & t'\\ t & r'\end{smallmatrix}\end{bmatrix}$
with
$r=\begin{bmatrix}\begin{smallmatrix}r_{xx} & r_{xy}\\ -r_{xy} & r_{yy}\end{smallmatrix}\end{bmatrix}$
and
$t=\begin{bmatrix}\begin{smallmatrix}t_{xx} & t_{xy}\\ -t_{xy} & t_{yy}\end{smallmatrix}\end{bmatrix}$
(and similarly for $r'$ and $t'$) \cite{Tse2011}.
We therefore have
\begin{align}
\begin{bmatrix}\bm{E}^{ri} \\ \bm{E}^{tj}\end{bmatrix}
=
\begin{bmatrix}r & t'\\ t & r'\end{bmatrix}
\begin{bmatrix}\bm{E}^{0} \\ 0\end{bmatrix}
=
\begin{bmatrix}r\bm{E}^{0} \\ t\bm{E}^{0}\end{bmatrix}.
\end{align}
The explicit forms of $r$ and $t$ can be obtained by solving the boundary condition equations~(\ref{Boundary-conditions}) with the use of Faraday's law, i.e., $\nabla\times\bm{E}=-(1/c)\partial \bm{B}/\partial t$, in media $i$ and $j$.

The Faraday rotation angle is calculated from the arguments of the transmitted electric field (in medium $j$):
\begin{align}
\phi_{\mathrm{F}}(\Omega_0)=\left[\mathrm{arg}(E^t_+)-\mathrm{arg}(E^t_-)\right]/2,
\end{align}
where $E^t_\pm=E^t_x\pm i E^t_y$ are the left-handed ($+$) and right-handed ($-$) circularly polarized components of the transmitted electric field $\bm{E}^t$.
The explicit forms of the arguments $\mathrm{arg}(E^t_\pm)$ are given by
\begin{widetext}
\begin{align}
\mathrm{arg}(E^t_\pm)=\tan ^{-1}\left[\frac{4 \pi  \alpha  \sigma_{xx}^{\mathrm{I}}(\Omega_0)\pm 4 \pi  \alpha  \sigma_{xy}^{\mathrm{R}}(\Omega_0)}{\sqrt{\epsilon_i/\mu_i}+\sqrt{\epsilon_j/\mu_j}+4 \pi  \alpha  \sigma_{xx}^{\mathrm{R}}(\Omega_0)-4 \pi  \alpha  \sigma_{xy}^{\mathrm{I}}(\Omega_0)}\right].
\label{arguments}
\end{align}
\end{widetext}
By substituting Eqs.~(\ref{sigma_xx}) and (\ref{sigma_xy}) into Eq.~(\ref{arguments}), we can obtain numerically the value of the Faraday rotation angle $\phi_{\mathrm{F}}(\Omega_0)$ at the light frequency $\Omega_0$ in a single TI surface.
\begin{figure*}[!t]
\centering
\includegraphics[width=2\columnwidth]{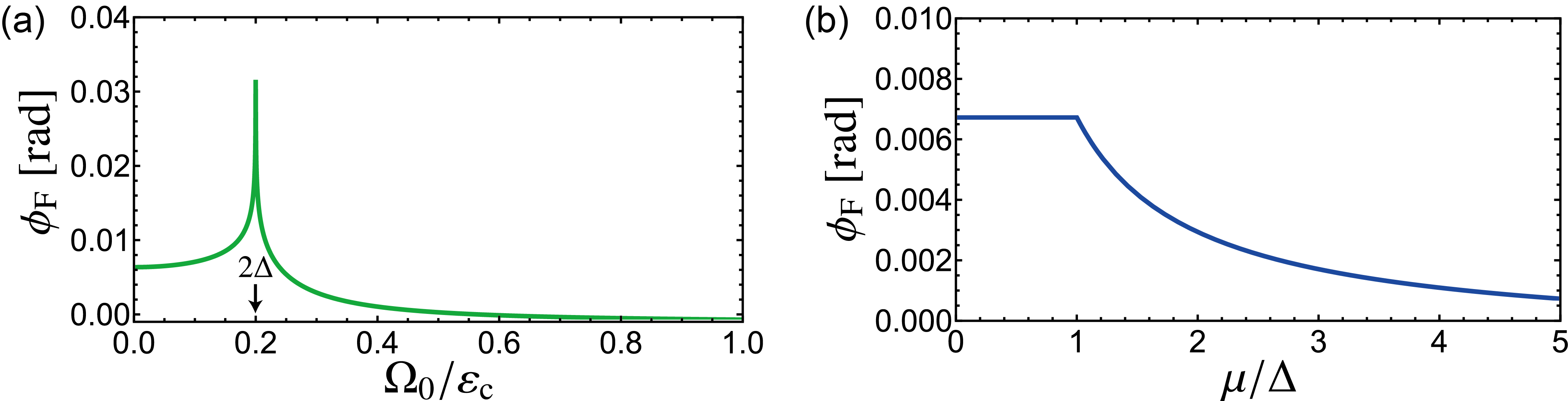}
\caption{(a) The input light frequency $\Omega_0$ dependence and (b) the chemical potential $\mu_{\mathrm{F}}$  dependence of the Faraday rotation angle $\phi_{\mathrm{F}}$ in the thin film regime.
We set $\varepsilon_{\mathrm{c}}=150\, \mathrm{meV}$ and $\Delta=15\, \mathrm{meV}$.
In (a) and (b), we also set $\mu/\Delta=0.5$ and $\Omega_0/\varepsilon_{\mathrm{c}}=0.05$, respectively.
}\label{Fig6}
\end{figure*}
\subsection{Faraday rotation angle in the thin film regime}
Here, we calculate the Faraday rotation angle in a TI thin film.
In the thin film regime where $d_{\mathrm{TI}}\ll \lambda=2\pi c/\Omega_0$, the thickness of the TI thin film can be neglected.
In other words, we may regard the system as a 2D system consisting only of the top and bottom surfaces.
Then, we can simply set $\sqrt{\epsilon_i/\mu_i}=\sqrt{\epsilon_j/\mu_j}\to 1$, as well as $\sigma_{xx}(\Omega_0)\to 2\sigma_{xx}(\Omega_0)$ and $\sigma_{xy}(\Omega_0)\to 2\sigma_{xy}(\Omega_0)$, which accounts for the contributions from both the top and bottom surfaces.

Figure~\ref{Fig6}(a) shows the input light frequency $\Omega_0$ dependence of the Faraday rotation angle $\phi_{\mathrm{F}}$, which reproduces the result in Ref.~\cite{Tse2010}.
The value of $\phi_{\mathrm{F}}$ in the low-frequency limit $\Omega_0/\varepsilon_{\mathrm{c}}\ll 1$ is almost universal such that $\phi_{\mathrm{F}}=\tan ^{-1}[\alpha(1-\Delta/\varepsilon_{\mathrm{c}})]\simeq \tan ^{-1}\alpha$.
Figure~\ref{Fig6}(b) shows the chemical potential $\mu_{\mathrm{F}}$ dependence of $\phi_{\mathrm{F}}$.
The value of $\phi_{\mathrm{F}}$ is constant as long as the chemical potential is in the surface gap (i.e., $\mu/\Delta<1$), reflecting the constant anomalous Hall conductivity $\sigma_{xy}^{\mathrm{R}}=(\alpha/4\pi)(1-\Delta/\varepsilon_{\mathrm{c}})$ in the topological transport regime.
On the other hand, we can see that the value of $\phi_{\mathrm{F}}$ begins to decrease as the carrier density, i,e., the value of $\mu/\Delta(>1)$, becomes larger.

\section{Transduction Efficiency in the Heterostructures Consisting of Magnetically Doped Topological Insulators and Nonmagnetic Insulators \label{Appendix-MTI}}
In this Appendix, we estimate the transduction efficiency in the heterostructures consisting of magnetically doped TIs (MTIs) and nonmagnetic insulators [see Fig.~\ref{Fig2}(a)].
To this end, let us characterize the spin density of the MTIs such as $X_{x}$(Bi$_{1-y}$Sb$_{y}$)$_{2-x}$Te$_{3}$ ($X$ = Cr or V) by the ratio to that of YIG,
\begin{align}
n_{\mathrm{s,MTI}}/n_{\mathrm{s,YIG}},
\end{align}
where $n_{\mathrm{s,YIG}}= 2.1\times 10^{19}\mu_{\mathrm B}\, \mathrm{mm}^{-3}$ (which is used in the main text).
We assume that $n_{\mathrm{s,MTI}}/n_{\mathrm{s,YIG}}<1$, because the composition ratio of the magnetic dopant is small like Cr$_{0.15}$(Bi$_{0.15}$Sb$_{0.85}$)$_{1.85}$Te$_{3}$ and V$_{0.15}$(Bi$_{0.2}$Sb$_{0.8}$)$_{1.85}$Te$_{3}$ \cite{Tokura2019,Liu2023}.
We also characterize the decay rate of the magnons in the ferromagnetic resonance state of TMIs by the ratio to that of YIG,
\begin{align}
\gamma_{\mathrm{MTI}}/\gamma_{\mathrm{YIG}},
\end{align}
where $\gamma_{\mathrm{YIG}}/2\pi= 1\, \mathrm{MHz}$ (which is used in the main text).
We assume that $\gamma_{\mathrm{MTI}}/\gamma_{\mathrm{YIG}}>1$, because uniform magnetic doping is quite difficult, resulting in ``dirty'' magnetic properties compared to those of YIG.

\begin{figure}[!b]
\centering
\includegraphics[width=\columnwidth]{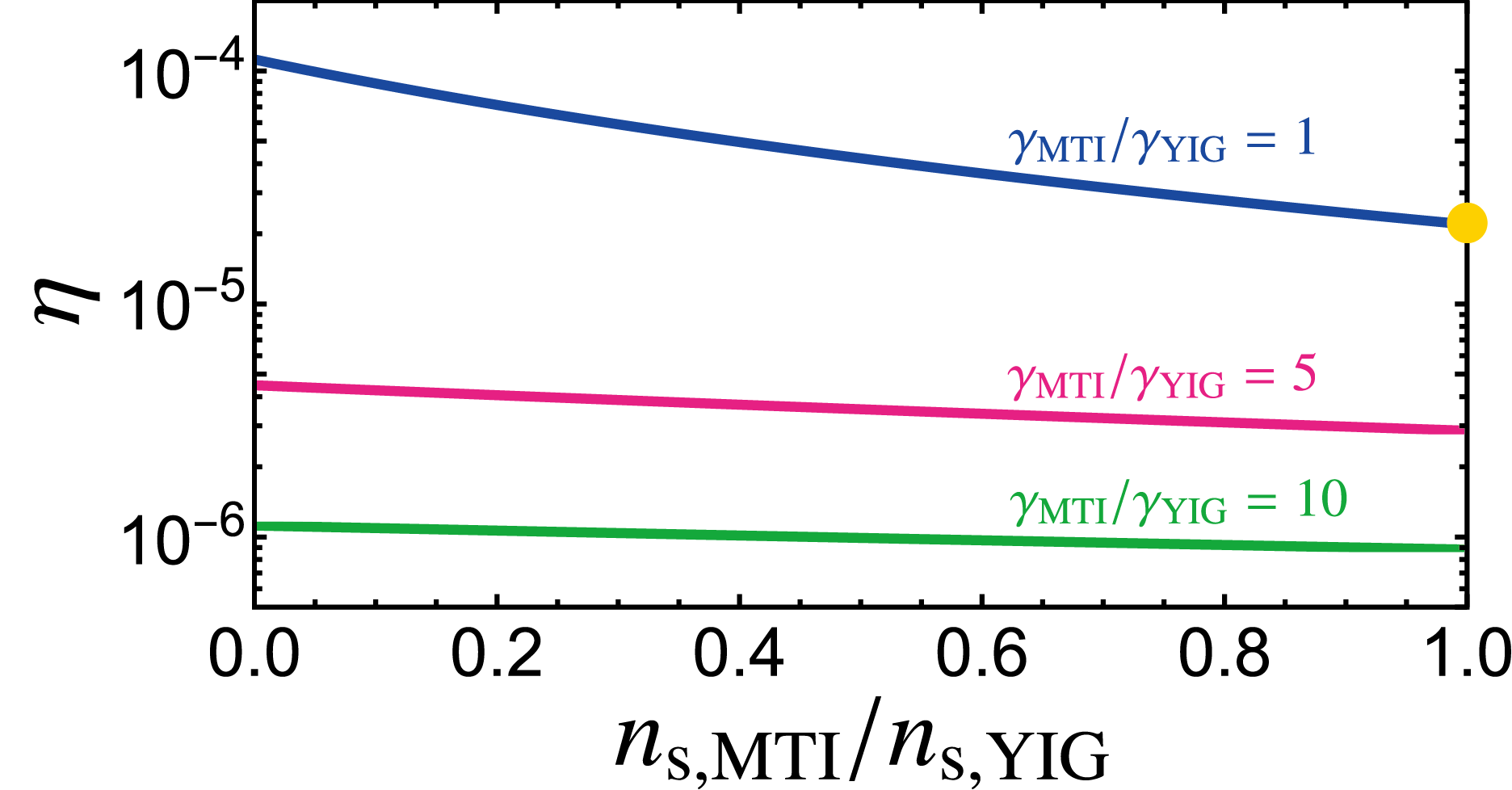}
\caption{Transduction efficiency of a heterostructure consisting of magnetically doped topological insulators (MTIs) and nonmagnetic insulators as functions of the spin density $n_{\mathrm{s,MTI}}$ and the magnon decay rate $\gamma_{\mathrm{MTI}}$ of MTIs.
We set $N_{\mathrm{L}}=30$ and $d_{\mathrm{MTI}}=5\, \mathrm{nm}$.
The other parameters are the same as in Fig.~\ref{Fig4} of the main text.
The value at $n_{\mathrm{s,MTI}}/n_{\mathrm{s,YIG}}=\gamma_{\mathrm{MTI}}/\gamma_{\mathrm{YIG}}=1$, highlighted by filled yellow circle, corresponds to the value in Fig.~\ref{Fig4} (the $\Omega_0\to 0$ limit of the blue line) of the main text.
}\label{Fig7}
\end{figure}
Figure~\ref{Fig7} shows the transduction efficiency $\eta$ of a heterostructure consisting of magnetically doped topological insulators (MTIs) and nonmagnetic insulators as functions of the spin density $n_{\mathrm{s,MTI}}$ and the magnon decay rate $\gamma_{\mathrm{MTI}}$.
In order to calculate the transduction efficiency, we can simply replace the FI layers and TI layers in the main text by the MTI layers and nonmagnetic insulator layers, respectively.
We see that the transduction efficiency gets improved as the spin density becomes low, which can be understood from the relation $\eta\sim \phi_{\mathrm{F}}^2/N_{\mathrm{s}}\sim 1/N_{\mathrm{s}}$ with $N_{\mathrm{s}}=n_{\mathrm{s,MTI}}V_{\mathrm{MTI}}$ because of the topological Faraday effect $ \phi_{\mathrm{F}}\approx  \phi_{\mathrm{F,TI}}=1/137$ (i.e., the contribution from the bulk of the MTI layers to the Faraday rotation angle can be neglected in the thin film limit).
Interestingly, the transduction efficiency converges in the low spin-density limit $n_{\mathrm{s,MTI}}/n_{\mathrm{s,YIG}}\to 0$.
This can be understood from that the $N_{\mathrm{s}}$ dependences of $\mathcal{C}\propto N_{\mathrm{s}}$ and $\zeta\propto 1/N_{\mathrm{s}}$ are canceled out by each other and $\mathcal{C}\to 0$ in the limit of $N_{\mathrm{s}}\to 0$ in Eq.~(\ref{Transduction-efficiency}) of the main text.
On the othe hand, the transduction efficiency is an decreasing function of $\gamma_{\mathrm{MTI}}$.
This can be understood from that the transduction efficiency behaves like $\eta\sim 1/\gamma_{\mathrm{MTI}}^2$ in Eq.~(\ref{Transduction-efficiency}).
To summarize these results, it is important to realize ``clean'' MTIs with low spin density and with low magnon decay rate.


\end{document}